\documentclass[a4paper]{jpconf}
\usepackage{graphicx}

\newcommand\beq{\begin{equation}}
\newcommand\eeq{\end{equation}}
\newcommand\bea{\begin{eqnarray}}
\newcommand\eea{\end{eqnarray}}
\newcommand\bI{\begin{itemize}}
\newcommand\eI{\end{itemize}}

\def\D{\Delta}

\def\D{{\mathcal{D}}}

\def\Z{{\bf z}}
\def\t{{\tau}}

\def\s{{\sigma}}
\def\a{\alpha}
\def\b{\beta}
\def\e{\epsilon}
\def\z{\xi}

\def\Tr{{\rm Tr}}

\newcommand\bra[1]{\left<#1\right|}
\newcommand\ket[1]{\left|#1\right>}

\def\half{\frac {1} {2}}

\begin{document}
\title{A review of the decoherent histories approach to the arrival time problem in quantum theory}

\author{James M Yearsley}

\address{Theoretical Physics Group, Blackett Laboratory, Imperial College, London SW7 2BZ, UK }

\ead{james.yearsley@imperial.ac.uk}

\begin{abstract}
We review recent progress in understanding the arrival time problem in quantum mechanics, from the point of view of the decoherent histories approach to quantum theory. We begin by discussing the arrival time problem, focussing in particular on the role of the probability current in the expected classical solution. After a brief introduction to decoherent histories we review the use of complex potentials in the construction of appropriate class operators. We then discuss the arrival time problem for a particle coupled to an environment, and review how the arrival time probability can be expressed in terms of a POVM in this case. We turn finally to the question of decoherence of the corresponding histories, and we show that this can be achieved for simple states in the case of a free particle, and for general states for a particle coupled to an environment.
\end{abstract}

\section{Introduction}

Questions involving time in quantum theory have a long and controversial history \cite{time}. Quantities such as arrival and dwell times, despite being measureable \cite{meas}, still lack concrete grounding within ``standard'' quantum theory. Arrival times, in particular, have attracted much interest, as the natural procedure of quantising the appropriate classical quantity gives rise to an operator which is not self-adjoint and thus, in standard quantum theory at least, cannot easily be considered as an observable. 

Despite these difficulties, if one considers a free particle
in an initial state $\rho= \ket{\psi}\bra{\psi} $
localized in $x>0$ consisting entirely of negative momenta, and asks
for the probability $p(t_1,t_2)$ that the particle crosses the origin
during the time interval $[t_1,t_2]$ there {\it is} a semi-classical answer, given by the difference between the probability of being in $x>0$ at the initial and final time \cite{time}. Defining $P(t)=\theta(\hat x(t))$,
\bea
p (t_1, t_2)&=&\Tr(P(t_{1})\rho)-\Tr(P(t_{2})\rho)=-\int_{t_1}^{t_2} dt \Tr(P \dot \rho_{t})\nonumber\\
&=&\int_{t_{1}}^{t_{2}}dt J(t)
\label{1},\eea
where
\bea
J(t)&=&\frac{i}{2m}\left(\psi^{*}(0,t)\frac{\partial \psi(0,t)}{\partial x}-\frac{\partial \psi^{*}(0,t)}{\partial x}\psi(0,t)\right)\nonumber
\eea
is the standard quantum mechanical probability current. (We denote the state at time $t$ by $\rho_{t}$, and set $\hbar=1$ throughout.) This can also be written in the following two forms,
\bea
p(t_{1},t_{2})&=& \int_{t_{1}}^{t_{2}}dt \int dp dq \frac{(-p)\delta(q)}{m}W_{t}(p,q)\label{1w},
\eea
where $W_{t}(p,q)$ is the Wigner function, defined later, and
\bea
p(t_{1},t_{2}) &=& \Tr(C\rho)\nonumber\\
C &=& \int_{t_1}^{t_2} dt  \frac { (-1) } {2 m } \left( \hat p \delta ( \hat x(t) ) + \delta (\hat x(t) ) \hat p \right)\nonumber\\
&=& P ( t_1 ) - P (t_2 ).
\label{2}
\eea
These expressions agree with the classical result, with the Wigner function $W$ replaced by the classical distribution function $w$, provided that the classical trajectories are straight lines. They are also correctly normalized to 1
as $t_2 \rightarrow \infty $ with $t_1 = 0$, but they are not generally positive, even for states consisting entirely of negative momenta. This genuinely quantum phenomenon is called backflow
and arises because the operator $C$, positive classically, has negative
eigenvalues  \cite{cur,BrMe,back}. This means we cannot generally regard Eq.(\ref{1}) as defining an arrival time distribution. There is, in addition, a more fundamental problem with Eq.(\ref{1}), which is that probabilities in quantum theory should be expressible in the form \cite{NC}
\beq
p(\a)=\Tr(P_{\a}\rho).\nonumber
\eeq
Here $P_{\a}$ is a projection operator, or more generally a POVM, associated with the outcome $\a$. Eq.(\ref{1}) cannot be expressed in this form, and we therefore conclude that it is not a fundamental expression in quantum theory, so must be the result of some approximation.

One interesting question is whether heuristic expressions for the arrival time probabilities of the form Eqs.(\ref{1}, \ref{1w}) can be derived in some limit from a more axiomatic approach. In this contribution we will show how this may be done in the decoherent histories approach to quantum theory. This contribution is based on a series recent papers \cite{HaYe1, HaYe2, HaYe3, jmy2}, but also draws on earlier work by Hartle, Halliwell and others \cite{more}.

Another issue is the relationship between the various time scales present in the problem. How long does it take a wavepacket to cross the origin? Is there a minimum time scale on which we can meaningfully talk about probabilities for crossing? One of the aims of this contribution is to clarify some of these issues.

The rest of this paper is structured as follows. In section \ref{sec6} we briefly review the decoherent histories approach to quantum theory in general, and the arrival time problem in particular. In section \ref{sec7} we examine how complex potentials may be used to define class operators for the arrival time problem. In section \ref{sec8} we then discuss the arrival time problem for open quantum systems, reviewing how environmentally induced decoherence allows the arrival probabilities to be written in terms of a POVM. In section \ref{sec9} we then turn to the question of whether the corresponding histories decohere. We summarise and discuss some open questions on section \ref{sec10}.

\section{The decoherent histories approach to the arrival time problem} \label{sec6}

\subsection{The Decoherent Histories approach to Quantum Theory}
We begin by briefly reviewing the decoherent histories approach to quantum theory.  More extensive discussions can be found in the references \cite{GeH1}. 

Alternatives at fixed moments of time in quantum theory are
represented by a set of projection operators $\{ P_a \}$,
satisfying the conditions
\beq
\sum_a P_a = 1, \quad P_a P_b = \delta_{ab} P_a,\nonumber
\eeq 
where we take $a$ to run over some finite range. In the decoherent histories approach to quantum
theory histories are represented by class operators $C_{\a}$ which are time-ordered
strings of projections
\beq
C_{\a} = P_{a_n} (t_n) \cdots P_{a_1}(t_1)
\label{1.3}
\eeq
or sums of such strings  \cite{Ish}. Here the projections are in the Heisenberg
picture and $ \a $ denotes the string $ (a_1, \cdots a_n)$. 
All class operators satisfy the condition
\beq
\sum_{\a} C_{\a} = 1.
\label{1.4}
\eeq
Probabilities are assigned to histories via the formula
\beq
p(\a) = {\rm Tr} \left( C_{\a} \rho C_{\a}^{\dag} \right).
\label{1.6}
\eeq
However probabilities assigned in this way do not necessarily obey the
probability sum rules, because of quantum interference.
We therefore introduce the decoherence functional
\beq
D(\a, \b) = {\rm Tr} \left( C_{\a} \rho C_{\b}^{\dag} \right),\label{dec}
\eeq
which may be thought of as a measure of interference between pairs of histories.
We require that sets of histories satisfy the condition of
decoherence, which is
\beq
D(\a, \b) = 0, \ \ \ \a \ne \b\label{1.7}.
\eeq
This ensures that all probability sum rules are satisfied.

We note that when there is decoherence Eq.(\ref{1.4}) and Eq.(\ref{1.7}) together imply that the probabilities $p(\a)$ are given by the simpler expressions
\beq
q(\a) = {\rm Tr} \left( C_{\a} \rho \right).
\label{1.17}
\eeq
Decoherence ensures that $q(\a)$ is real and positive, even though it
is not in general. 
In this way decoherent histories may reproduce probabilities of the form Eq.(\ref{2}).

\subsection{The Decoherent Histories aproach to the arrival time problem}

We turn now to the definition of the arrival time problem in decoherent histories. It is natural to start by considering the class operator for not crossing the origin in a given time interval $[0,\t]$. Splitting this time into smaller intervals of size $\e$ and letting $\t=n\e$, the natural class operator for not crossing the origin during this time would be
\beq
C_{nc}=P(n\e)...P(2\e)P(\e)\label{cnc}.
\eeq 
This object represents the history where the state is in $x>0$ at the times $\e,2\e...$ but can be anywhere at intermediate times. An obvious thing to try is to take the limit as the spacing between projections approaches zero, with the hope of obtaining a precise specification of the behaviour of the state. However that this limit yields the restricted propagator.
\beq
\lim_{\e\to0}P(n\e)...P(2\e)P(\e)=g_{r}(\t,0).
\eeq

$g_{r}(\t,0)$ generates unitary evolution in the subspace of states with support only in $x>0$, so one find that the probability of crossing the origin is zero. This is a manifestation of the quantum Zeno effect \cite{Zeno}. The resolution of this problem is simple in principle, one simply declines to take the limit $\e\to0$ in Eq.(\ref{cnc}), and instead works with histories which have some finite ``resolution'' $\e$. In practice, however, the situation is complicated for two reasons. The first is that the object Eq.(\ref{cnc}) is difficult to work with analyticaly, although semi-classical approximations are avaliable \cite{HaYe2}. The second is that it is not clear what the limits are on the ``resolution'' $\e$. 

The solution to both of these problems is to show that the evolution represented by Eq.(\ref{cnc}) is equivalent to evolution in the presence of a simple complex potential $V(x)=-iV_{0}\theta(-x)$ \cite{HaYe3}. One can then use this equivalence either to derive the class operators for crossing and not crossing directly \cite{HaYe1}, or one can instead use it to deduce the minimum spacing allowed between the projections in Eq.(\ref{cnc}) \cite{HaYe3}. 

\section{The relationship between complex potentials and strings of projection operators}\label{sec7}

Probably the most significant development in the histories approach to time in quantum theory to have occured in the past few years has been the understanding of the relationship between class operators defined in the manner of Eq.(\ref{cnc}), and those defined via complex potentials. This development has been significant for three reasons,
\bI
\item It has made the calculation of arrival time probabilities within decoherent histories relatively straightforward.
\item It has led to a much better understanding of the relationship between the different timescales that occur in the problem, in particular the Zeno time.
\item It has suggested new connections between the histories approach and some of the various detector based models for the computation of arrival time probabilities.
\eI

The proof of this equivalence can be found in \cite{HaYe3}, building on earlier work by Echonabe et al \cite{Ech}. Here we will simply state the equivalence, and discuss some of the consequences for the arrival time problem.

The equivalence is expressed as follows,
\beq
e^{-iH\t}P(n\e)...P(2\e)P(\e)\ket{\psi}\approx e^{-iH\t-V_{0}\bar P\t}\ket{\psi}
\eeq
where
\beq
P=\theta(\hat x), \quad \bar P=\theta(-\hat x),\quad \mbox{and } n\e=\t\nonumber
\eeq
The equivalence holds provided 
\beq
V_{0}\e\sim1, \quad \mbox{and } \e<<\frac{1}{\Delta H}\nonumber
\eeq
Since we know from studies of the complex potential that we begin to have reflection when $V_{0}\sim E$, where $E$ is the energy of the initial state, it follows that the time scale on which the projections cause significant reflection is given by $\e\sim1/E$. There is therefore a regime,
\beq
\frac{1}{E}<<\e<<\frac{1}{\Delta H}
\eeq
in which evolution under strings of projection operators is equivalent to that under the complex potential {\it and} there is negligible reflection.

In Refs.\cite{HaYe1, HaYe2}  the class operators corresponding to a first crossing of the origin in the time interval $[t_{k-1}, t_{k}]$ were computed to be
\beq
C_k =  \bar P ( t_{k} ) P (t_{k-1}) \cdots P(t_2) P(t_1),
\label{24}
\eeq 
for $ k \ge 2 $, with $C_1 = \bar P (t_1)$. These clearly describe histories
which are in $ x>0$ at times $ t_1, t_2, \cdots t_{k-1}$ and in $x<0$ at
time $t_k$, so, approximately, describe a first crossing between $t_{k-1}$ and $t_k$. 
These class operators can also be expressed in terms of evolution under the complex potential \cite{HaYe1},
\bea
C_{k}&=&\int_{t_{k}}^{t_{k+1}}dt C(t)\label{ccp}\\
 C(t)&=&e^{-iH(\t-t)}Ve^{-iHt-Vt}
\eea
The object $C(t)$ is the class operator for crossing at a time $t$. However this designation is somewhat misleading, because any state that enters within a time $\sim1/V_{0}$ of $t$ will also contribute to this crossing probability, because of the finite time it takes the complex potential to absorb the state. Differently put, the ``detector'' modeled by the complex potential has a time resolution of order $\sim1/V_{0}$, and thus we cannot  use it to define arrival times more accurately than this.

Given this, it is natural to use class operators like Eq.(\ref{ccp}) to define arrival time. If the time interval is much  greater than $1/V_{0}$ the dependence on $V_{0}$ drops out and we can show that \cite{HaYe1}
\bea
e^{iH\t}C_{k}&=&\int_{t_{k}}^{t_{k+1}}dt e^{i H t}Ve^{-iHt-Vt}\nonumber\\
&\approx&P(t_{k})-P(t_{k+1})\label{csp}, \quad \mbox{providing } t_{k+1}-t_{k}>>1/V_{0}
\eea
In Ref \cite{HaYe2} Eq.(\ref{csp}) was also derived from Eq.(\ref{24})  with the help of a simple semiclassical approximation \footnote{The situation is more subtle than implied in the derivation in \cite{HaYe1, HaYe2}. The class operators Eq.(\ref{cnc}, \ref{ccp}) are the quantum analogue of {\it first} crossing operators, they are non-zero only for states incoming from the right, and they are non-negative for any initial state. By contrast, the class operators Eq.(\ref{csp}) are non-zero for states incoming from the left, and can be negative for states incoming from the right. The first of these difficulties can be handled straightforwardly \cite{jmy3}, but the second, a manifestation of backflow, requires conditions on the state. Fortunately it can be shown that states exhibiting decoherence of arrival time probabilities will not display backflow \cite{HaYe1}.}. In fact, in \cite{jmy2} it was shown that this semi-classical approximation is closely related to the decoherence conditions discussed below.

In the case of a free particle without an environment, this class operator reproduces Eq.(\ref{1}) under the conditions of decoherence. This is because, assuming decoherence, the arrival time probability is given by
\beq
p(t_{2},t_{1})=\Tr(C\rho C^{\dagger})=\Tr(C \rho),
\eeq
using Eq.(\ref{1.17}). This is an important result, we have succeeded in deriving probabilities of the form Eq.(\ref{1}) from decoherent histories, in the limit that the corresponding histories decohere.

\section{Arrival time for open systems}\label{sec8}

We will see below that decoherence of arrival time histories can only be achieved for a generic initial state in the presence of an environment. It is therefore important to look more generally at the arrival time problem for the case of a state coupled to an environment. There are two aspects to consider. The first is, given that we expect the arrival time probabilities to be given in terms of the current, what are the relavent properties of the current for a system coupled to an environment? The second issue is to examine how the inclusion of an environment helps to produce decoherence of the corresponding arrival time histories. We will deal with the first issue in this section, and put off the discussion of decoherence till later.

The arrival time probability $p(t_{1},t_{2})$ can be written in terms of the Wigner function, as
\bea
p(t_{1},t_{2})&=& \int_{t_{1}}^{t_{2}}dt \int dp dq \frac{(-p)\delta(q)}{m}W_{t}(p,q)\label{2w},
\eea
where the Wigner function is defined as
\beq
W(p,q)=\frac{1}{2\pi}\int d\z \rho(q+\z/2,q-\z/2)e^{-i p \z}\nonumber
\eeq
The difficulty with Eq.(\ref{2w}) is that for a general state $W(p,q)$ need not be positive, it is therefore possible that this ``probability'' could be negative \footnote{Note that this negativity of the current had nothing to do with the possibility of having momentum components with both signs in the initial state. We have chosen our initial state to have only negative momenta. Clearly Eq.(\ref{2w}) will also have to be modified for the case of an initial state with momenta of both signs, but this is a separate issue that occurs equally in the classical case, and it can be dealt with in a reasonably straightforward manner, \cite{time, jmy3} }. 
An obvious solution to this problem is to consider a state coupled to an environment. This is because it is known that such evolution causes the Wigner function of the system to become positive on a very short timescale. After this time, an expression like Eq.(\ref{2w}) may be an acceptable, if still heuristic, arrival time distribution for a system coupled to an environment.

In \cite{jmy2} this scheme was examined, using as an environment the model of quantum brownian motion \cite{CaLe}. The model can be expressed in terms of a master equation, in general quite complicated, but which simplifies in the limit of high temperatures and negligible dissipation to the following \cite{jmy2},
\beq
\frac{\partial \rho_{t}}{\partial t}(x,y)=\frac{i}{2m}\left(\frac{\partial^{2}}{\partial x^{2}}-\frac{\partial^{2}}{\partial y^{2}}\right)\rho_{t}(x,y)-D(x-y)^{2}\rho_{t}(x,y)\nonumber
\eeq
A further interesting limit of this evolution is the so called ``near-deterministic limit'' \cite{jmy2} of $t<<\t_{s}=p_{0}^{2}/D$. It is easy to show that $\t_{s}$ is the time scale on which an initial state sharply peaked around momentum $p_{0}<0$ begins to have significant probability to be found with positive momentum. For times much less than this, although the evolving state will have a finite momentum width it will still approximately follow the deterministic classical trajectory. This is obviously an interesting limit for the arrival time problem. 
It was found that, in this limit, Eq.(\ref{2w}) defines an acceptable arrival time distribution for a generic initial state, after a time
\beq
t=\left(\frac{3}{16}\right)^{1/4} \left(\frac{2m}{D}\right)^{\half}=\left(\frac{3}{16}\right)^{1/4}\t_{l}
\eeq

Introducing the notation,
\beq
{\bf z}=\left(\begin{array}{r}p \\ q\end{array}\right)=\left(\begin{array}{c}z_{0}\\z_{1}\end{array}\right)\nonumber
\eeq
and the class of Gaussian phase space functions,
\beq
g(\Z;A)=\frac{1}{2\pi |A|^{\half}}\exp\left(-\half \Z^{T}A^{-1}\Z\right)\nonumber
\eeq
After this time we may write the current as
\beq
J(t)=\Tr(F \rho)
\eeq
where
\beq
F=\int d\Z P_{\Z}\frac{(-z_{0})}{m}\delta(z_{0}+z_{1}t/m)
\eeq
and we have defined the POVM
\beq
P_{\Z}=\frac{1}{\pi}\int d\Z' \ket{\Z'}\bra{\Z'}g(\Z-\Z';B),\quad \mbox{with } B=\left(\begin{array}{cc}2Dt-s^{2}&Dt^{2}/m\\
Dt^{2}/m& 2Dt^{3}/3m^{2}-1/4s^{2}\end{array}\right)
\eeq

We may now go further, and write the arrival time probability as,
\beq
p(t_{1},t_{2})=\int_{t_{1}}^{t_{2}}dt \Tr(F \rho)\approx\Tr(E\rho)\label{arvp}
\eeq
where
\beq
E=\int d\Z P_{\Z}[\theta(z_{1}+z_{0}t_{1}/m)-\theta(z_{1}+z_{0}t_{2}/m)]
\eeq
is a POVM representing arrival at $x=0$ between $t_{1}$ and $t_{2}$. This operator is just a smeared version of the classical phase space operator, but crucially it is positive for $p<0$. Since the decoherent histories approach to the arrival time problem gives the current as the correct arrival time distribution when there is decoherence, Eq.(\ref{arvp}) gives the arrival time probabilities  computed via decoherent histories.

The important point here is that we have managed to write the arrival time probabilities Eq.(\ref{2w}), in terms of the expectation value of a POVM, thus bringing them into line with other expressions for probabilities in quantum theory.

\section{Decoherence of arrival time probabilities}\label{sec9}

We turn now to discussing the conditions under which we have decoherence of crossing probabilities.  The general picture we have in mind is  an initial wavepacket defined at $t=0$, evolved until time $t_{1}$, possibly in the presence of an environment , and then we wish to compute the probability of crossing between $t_{1}$ and $t_{2}$.

Consider first the decoherence functional for the histories correspnding to crossing in the intervals $[t_{k},t_{k+1}]$ and $[t_{j},t_{j+1}]$, where without loss of generality we take $t_{j+1}\leq t_{k}$
\bea
D_{kj}&=&\bra{\psi}(P(t_{k})-P(t_{k+1}))(P(t_{j})-P(t_{j+1}))\ket{\psi}\nonumber\\
&=&\bra{\psi}(P(t_{k})- P(t_{k+1}))(\bar P(t_{j+1})-\bar P(t_{j}))\ket{\psi}\nonumber
\eea
It is a sum of terms of the form
\beq
d_{kj}=\bra{\psi}P(t_{k})\bar P(t_{j})\ket{\psi}\nonumber
\eeq
with $t_{j}\leq t_{k}$. Note that
\beq
|d_{kj}|^{2}\leq\Delta_{kj}=\bra{\psi} P(t_{k}) \bar P(t_{j}) P(t_{k})\ket{\psi}\label{delta}
\eeq
The key point is that $\Delta$ has the form of a probability, it is essentially the probability to find the state in $x<0$ at $t_{j}$ and then in $x>0$ at $t_{k}$. Since we will be considering left moving wavepackets, this probability should be small. To show we have decoherence it therefore suffices to compute the quantity $\Delta_{kj}$ defined in Eq.(\ref{delta}). We will do this both for a free particle and a particle coupled to an environment. If we include an environment we anticipate that this will be small simply from the form of Eq.(\ref{delta}). This is because the effect of the environment is to cause the density matrix to become tightly peaked around the classical path and classically for $p<0$ the probability given by $\Delta_{kj}$ is zero. We will see how this works in a specific example below.

We take $k=1, m=2$ without loss of generality, and we drop the subscript from now on
\beq
\Delta=\Tr(P(t_{2})\bar P(t_{1})\rho \bar P(t_{1}))=\int_{\a}\D x \int_{\a} \D y \exp\left(i S[x]-i S[y]- D\int dt(x-y)^{2}\right) \rho_{t_{1}}(x_{1},y_{1}),
\eeq
where the histories $\a$ are those that start at $x_{1},y_{1}<0$ and finish at $x_{2}>0$. In terms of the density matrix propagator \cite{HaZo}
\bea
\Delta&=&\int_{0}^{\infty}dx_{2}\int_{-\infty}^{0}dx_{1}\int_{-\infty}^{0}dy_{1}J(x_{2},x_{2},t_{2}|,x_{1},y_{1},t_{1})\rho_{t_{1}}(x_{1},y_{1})\nonumber \\
&=& \int_{0}^{\infty}dx_{2}\int_{-\infty}^{0}dx_{1}\int_{-\infty}^{0}dy_{1} \left(\frac{m}{\pi t}\right)\nonumber \\ &&\exp\left(\frac{im}{2(t_{2}-t_{1})}((x_{2}-x_{1})^{2}-(x_{2}-y_{1})^{2})-\frac{D(t_{2}-t_{1})}{3}(x_{1}-y_{1})^{2}\right)\rho_{t_{1}}(x_{1},y_{1})\nonumber.
\eea
Transforming to new variables
\beq
X=\frac{x_{1}+y_{1}}{2},\quad \z=x_{1}-y_{1},
\eeq
and writing the density matrix in terms of the Wigner function via
\beq
\rho(x,y)=\int_{-\infty}^{\infty}dp e^{i p(x-y)}W(p,\frac{x+y}{2}),
\eeq
we obtain
\bea
\Delta&=&-\int_{0}^{\infty}dx_{2}\int_{-\infty}^{0}dX\int_{-X}^{X}d\z \int_{-\infty}^{\infty}dp \left(\frac{m}{2\pi (t_{2}-t_{1})}\right) \nonumber \\ 
&&\exp\left(\frac{im}{(t_{2}-t_{1})}\z(X-x_{2})+i p\z-\frac{D(t_{2}-t_{1})}{3}\z^{2}\right)W_{t_{1}}(X,p)\label{int}.
\eea
We see from this expression that there is a time scale, $\t_{l}=\sqrt{2m/D}$, set by the environment on which localisation effect are important. This timescale is the same as that on which the current becomes positive, as we saw earier.

There are three cases to explore here, the first is where there is no environment, $D=0$. This is the case covered in Ref.\cite{HaYe1}. 
The second case is the intermediate one, $t_{1}/\t_{l}>>1$ but $(t_{2}-t_{1})/\t_{l}<<1$. This is the most general case in which we can expect to have environmentally induced decoherence.
The final case is where $t_{1}/\t_{l}, (t_{2}-t_{1})/\t_{l}>>1$. This is the case of very strong environmental coupling.  

The details of the calculation of $\Delta$ in each case are given in \cite{jmy2}, and also for the free particle case in \cite{HaYe1}, so we simply quote the result in each case.

\subsection{Free particle case}

We assume that our state at $t_{1}$ is of the form
\beq
W_{t_{1}}(X,p)=\frac{1}{\pi}\exp\left(-\frac{(X-X_{0}-p_{0}t_{1}/m)^{2}}{2\s^{2}}-2\s^{2}(p-p_{0})^{2}\right),
\eeq
and that $\s^{2}$ is large, so the state is tightly peaked in momentum. After some calculation, and assuming $p_{0}^{2}(t_{2}-t_{1})/2m>>1$, we obtain
\beq
\Delta < \frac{1}{\s |p_{0}|}<<1,
\eeq
there will therefore be decoherence for gaussian wavepackets tightly peaked in momentum, ie $|p_{0}|\sigma>>1$, provided $E_{0}(t_{2}-t_{1})>>1$. In \cite{HaYe1} it was argued that this conclusion also holds for orthogonal superpositions of gaussians.

\subsection{Weak environment}

Turning to the intermediate case, although we can still neglect the effects of the environment for the time it takes to cross the origin, the key point is that $W_{t_{1}}(X,P)$ is the initial state evolved with an environment, and since $t_{1}/\t_{l}>>1$ this should be significant. Because the Wigner function propagator is a gaussian the analysis is similar to the first case. We find, after some calculation, 
\beq
\Delta<\sqrt{\frac{1}{E_{0}t_{1}}}\left(\frac{\t_{l}}{t_{1}}\right)<<1.
\eeq

\subsection{Strong environment}
Finally we have the case of strong system-environment coupling. In this case we can show that $\Delta$ will be very small, provided $(t_{2}-t_{1})<<\t_{s}$. This gives us an {\it upper} bound on the time interval, $[t_{1},t_{2}]$, rather than a lower one. The lower time scale is provided by the condition $t_{2}-t_{1}>>\t_{l}=\sqrt{2m/D}$. This lower time scale is compatible with the condition that the current be positive.

Note however that this lower limit is state independent. There will be states for which arrival time probabilities decohere on much shorter time scales than this, for example the simple cases which decohere in the absence of any environment will continue to do so in the presence of an environment, at least until a time $\sim \t_{s}$.

\subsection{Discussion}

What one draws from studying these cases is that whether or not one can assign arrival time probabilities in decoherent histories depends on the form of the state at the time it crosses the origin.  Environmentally induced decoherence produces mixtures of gaussian states from generic initial ones, and thus after a short time arrival time probabilities can be defined whatever the initial state. However it is not necessary for the system to be monitored whilst it is crossing the origin. The smallest time interval,  $\delta t$, for which we can define an arrival time probability is therefore set by the energy of the system and not the details of the environment and we must have $E \delta t>>1$. This is in agreement with Ref.\cite{HaYe1}, and also with the results of earlier works, concerning the accuracy with which a quantum system may be used as a clock \cite{clock}.

In conclusion, for a general state, decoherence of histories requires that we evolve for a time much greater than $\t_{l}=\sqrt{2m/D}$ before the first crossing time. This is because $\t_{l}$ is the time scale on which quantum correlations disappear and our  state begins to resemble a mixture of gaussian states. After this time, we may define arrival time probabilities to an accuracy $\delta t$, provided only that $E\delta t>>1$. States which start as gaussians may be assigned arrival time probabilities without this initial period of evolution. This is in line with the general result that some coarse-graining is always required to achieve a decoherent set of histories in quantum theory.

\section{Summary and open questions}\label{sec10}

In this contribution we have given an overview of the way in which decoherent histories  gives insight into the arrival time problem in quantum theory.  The key steps are as follows, in section \ref{sec6} we discussed how probabilities are defined in decoherent histories, and how they can reduce to the form Eq.(\ref{1.17}) when there is decoherence. Next in section \ref{sec7} we saw how the class operators can be derived using complex potentials, and how they reduce to the form Eq.(\ref{csp}) under certain conditions. Then in section \ref{sec8} we showed that the same is true for a particle coupled to an environment, under certian conditions. Finally in section \ref{sec9} we proved that we have decoherence of histories for an interesting class of states for a free particle, and for a generic state for a particle coupled to an environment. Although space has prevented us from giving all the details, we hope that the general picture is clear.

Although the arrival time probabilities computed from decoherent histories agree with the heuristic ones when we have decoherence, their derivation in this way represents a significant advance in our understanding. The difficulty with regarding the current as the arrival time distribution is the arbitrary way in which one accepts these ``probabilities'' when they are positive, but declines to do so when they are not. Because decoherence is an essential part of the histories formalism, this arbitrariness is replaced with a consistent set of rules governing when probabilities may or may not be assigned. Whilst this may be of no consequence in the setting of a laboratory, it may prove hugely important in the analysis of closed systems, in particular the study of quantum cosmology \cite{jjhqc}. 

Although the arrival time problem is now well understood from this perspective, there are a number of other areas, such as dwell or tunneling times, where the methods outlined here could be usefully applied. 
Finally, we mentioned right at the beginning of this contribution that there are a multitiude of different approaches to tackling the arrival time problem. An interesting question is how the decoherent histories approach is related to other approaches, particularly to those operational approaches that rely on coupling to model clocks. Further study of this issue may shed light on the reasons why there seem to be so many divergent approaches to the arrival time problem.

\ack I would like to thank the organisers of DICE 2010, especially Hans-Thomas Elze, for giving me the opportunity to take part in this conference. The majority of the work reported here was carried out in collaboration with Jonathan J Halliwell, to whom I am also indebted for many years of support and stimulating discussion.


\section*{References}
\begin{thebibliography}{10}



\bibitem{time} See for example; J.G.Muga, R.Sala Mayato and I.L.Egusquiza (eds),
{\it Time in Quantum Mechanics} (Springer, Berlin, 2002);  J.G. Muga, A. Ruschhaupt and A. del Campo (eds), {\it Time in Quantum Mechanics - Vol. 2} (Springer, Berlin, 2010); J.G.Muga and C.R.Leavens, Phys.Rep. 338, 353 (2000).

\bibitem{meas} J.A. Damborenea, I.L. Egusquiza, G.C. Hegerfeldt, and J.G. Muga,
Phys. Rev. A 66, 052104 (2002); 
G. C. Hegerfeldt, D. Seidel, J. G. Muga, Phys. Rev. A 68, 022111 (2003).

\bibitem{cur} J. G. Muga, J. P. Palao and C. R. Leavens
Phys. Lett. A 253, 21 (1999).

\bibitem{BrMe} A.J.Bracken and G.F.Melloy, J.Phys. A27, 2197
(1994).

\bibitem{back} S.P.Eveson, C.J.Fewster and R.Verch,
Ann.Inst. H.Poincar\'e 6, 1 (2005); M.Penz, G.Gr\"ubl, S.Kreidl and P.Wagner,
J.Phys. A39, 423 (2006).

\bibitem{NC} See for example; M.A.Nielsen and I.L.Chuang, {\it Quantum computation and quantum information} (Cambridge University Press, Cambridge, UK 2000); P.Busch, M.Grabowski and P.Lahti, {\it Operational Quantum Physics} (Springer-Verlag, Berlin 1997).

\bibitem{HaYe1} J.J.Halliwell and J.M.Yearsley, Phys. Rev. A 79, 062101 (2009).


\bibitem{HaYe2}  J.J.Halliwell and J.M.Yearsley, Phys. Lett. A 374, 154 (2009)

\bibitem{HaYe3} J.J.Halliwell and J.M.Yearsley, J.Phys. A 43, 445303 (2010).

\bibitem{jmy2} J.M.Yearsley, Phys. Rev. A 82, 012116 (2010).

\bibitem{more} See for example,  J. B. Hartle, Phys. Rev. D 44, 3173 (1991); J.J.Halliwell and E.Zafiris, {Phys.Rev.} { D57}, 3351 (1998); R.J.Micanek and J.B.Hartle, {Phys.Rev.} { A54}, 3795 (1996);
 N.Yamada and S.Takagi, {Prog.Theor.Phys.}
{ 85}, 985 (1991); { 86}, 599 (1991); { 87}, 77 (1992); N. Yamada,
{ Sci. Rep. T\^ohoku Uni., Series 8}, { 12}, 177 (1992).

\bibitem{GeH1} M.Gell-Mann and J.B.Hartle,
in {\it Complexity, Entropy and the Physics of Information, SFI
Studies in the Sciences of Complexity}, Vol. VIII, W. Zurek (ed.)
(Addison Wesley, Reading, 1990); {Phys.Rev.} { D47}, 3345 (1993);  R.B.Griffiths, {J.Stat.Phys.} { 36}, 219 (1984);
{Phys.Rev.Lett.} { 70}, 2201 (1993); {Am.J.Phys.} { 55}, 11
(1987); R.Omn\`es, {J.Stat.Phys.} { 53}, 893 (1988);
{ 53}, 933 (1988); { 53}, 957 (1988); { 57}, 357 (1989); { 62},
841 (1991); {Ann.Phys.} { 201}, 354 (1990); {Rev.Mod.Phys.} { 64},
339 (1992); {\it The Interpretation of Quantum Mechanics}
(Princeton University Press, Princeton, 1994);  J.J.Halliwell,
in {\it Fundamental Problems in Quantum Theory}, edited by
D.Greenberger and A.Zeilinger, Annals of the New York Academy of
Sciences, Vol 775, 726 (1994); H.F.Dowker and J.J.Halliwell, {Phys. Rev.} {D46}, 1580 (1992).

\bibitem{Ish} C.Isham, J.Math.Phys. 35, 2157 (1994).

\bibitem{Zeno} B.Misra and E.C.G.Sudarshan, J.Math.Phys. 18, 756 (1977);  
A. Peres, Am. J. Phys. 48, 931 (1980).

\bibitem{Ech} J.Echanobe, A. del Campo and J.G.Muga, Phys. Rev. A 77, 032112
(2008).

\bibitem{jmy3} J.M.Yearsley (unpublished).

\bibitem{CaLe} A.O.Caldeira and A.J.Leggett, Physica A, 121, 587 (1983).



\bibitem{HaZo} J.J.Halliwell and A.Zoupas, Phys. Rev. D, 55, 4697 (1997).

\bibitem{clock} A.Peres, Quantum Theory: Concepts and Methods (Kluwer Academic, Dordrecht, 1993);
Y.Aharonov, J.Oppenheim, S.Popescu, B.Reznik and W.G.Unruh, Phys. Rev. A 57, 4130 (1998).

\bibitem{jjhqc} J.J.Halliwell, Phys. Rev. D, 80, 124032 (2009).





































\end{thebibliography}
\end{document}